N. K. Timofeyuk · D. Baye

# Hyperspherical Harmonics Expansion on Lagrange Meshes for Bosonic Systems in One Dimension

**Abstract** A one-dimensional system of bosons interacting with contact and single-Gaussian forces is studied with an expansion in hyperspherical harmonics. The hyperradial potentials are calculated using the link between the hyperspherical harmonics and the single-particle harmonic-oscillator basis while the coupled hyperradial equations are solved with the Lagrange-mesh method. Extensions of this method are proposed to achieve good convergence with small numbers of mesh points for any truncation of hypermomentum. The convergence with hypermomentum strongly depends on the range of the two-body forces: it is very good for large ranges but deteriorates as the range decreases, being the worst for the contact interaction. In all cases, the lowest-order energy is within 4.5% of the exact solution and shows the correct cubic asymptotic behaviour at large boson numbers. Details of the convergence studies are presented for 3, 5, 20 and 100 bosons. A special treatment for three bosons was found to be necessary. For single-Gaussian interactions, the convergence rate improves with increasing boson number, similar to what happens in the case of three-dimensional systems of bosons.

## 1 Introduction

The expansion in hyperspherical harmonics (HH) of a many-body wave function is a powerful scheme that allows the Schrödinger equation to be solved in a systematic way by reducing it to a system of coupled equations of only one dynamic variable for any number of particles $N$ [1]. This method is a standard one to describe few-body physics in three dimensions. The hyperspherical expansion in one dimension for a three-body problem has been considered in Refs. [2,3] for contact two-body potentials. Reference [2] also comments on the validity of the lowest-order approximation of the HH expansion at large $N$. The authors point out that the $N$-dependence of the lowest-order energy is quadratic while the exact solution gives a cubic $N$-dependence. This seems to contradict more recent observations made for three-dimensional systems that the HH expansion for a single-Gaussian two-body potential converges better with larger $N$ [4,5]. This means that the lowest-order energy should get closer to the exact solution with increasing $N$, which was checked numerically in Ref. [5]. Since the contact interaction can be seen as a limit of a single-Gaussian potential with a very short range, the same argument should hold for the contact interaction as well. Therefore, the statement of Ref. [2] regarding the quadratic versus cubic dependence of the lowest-order HH and exact energies needs to be reconsidered.

N. K. Timofeyuk (✉)
Physics Department, University of Surrey, Guildford, Surrey GU2 7XH, England, UK
E-mail: n.timofeyuk@surrey.ac.uk

D. Baye
Physique Quantique and Physique Nucléaire Théorique et Physique Mathématique, C.P. 229, Université Libre de Bruxelles (ULB), 1050 Brussels, Belgium
E-mail: dbaye@ulb.ac.be

In this paper, the HH expansion in one dimension is revisited and formulated in terms of an expansion on a single-particle oscillator basis [6,7]. The HHs are expressed as functions of symmetrized oscillator states. The matrix elements of the potential between these HHs are obtained from integral transforms for $N > 3$. The resulting coupled hyperradial equations depend on the hyperspherical quantum number $K$ and must be truncated at some value $K_{\max}$. For a given truncation $K_{\max}$, they are solved with the Lagrange-mesh method, an approximate variational technique which has the simplicity of a mesh calculation [8–10]. For a number of applications, this method provides accurate results with small numbers of basis functions [10]. Its accuracy is equivalent to the accuracy of a variational calculation with the same basis [11]. This property is in particular verified for HH equations in Refs. [10,12].

The convergence of the HH method with respect to $K$ is explored for the contact interaction, for which the exact solution is known, and for different ranges of a single-Gaussian interaction. The validity of extrapolation techniques is analysed.

In Sect. 2, the method of HH expansion is described for the one-dimensional case. The Lagrange-mesh method and an extension of this method are applied to the resulting hyperradial equations in Sect. 3. In Sect. 4, the special case of the lowest-order energy is considered and the convergence in the case of a contact interaction is analysed by comparison with the exact results. The same study is performed for Gaussian interactions in Sect. 5. Sect. 6 contains a summary while the Appendix provides information about some mathematical details relevant to the HH method.

## 2 Hyperspherical Harmonics Expansion and Schrödinger Equation in One Dimension

2.1 Choice of Coordinates

An $N$-body system is described by a set of $N - 1$ normalised Jacobi coordinates $\xi_1, \xi_2, \ldots, \xi_{N-1}$ defined in one dimension,

$$\xi_i = \sqrt{\frac{i}{i+1}} \left( \frac{1}{i} \sum_{j=1}^{i} r_j - r_{i+1} \right), \tag{1}$$

where $r_i$ is the individual coordinate of the $i$-th body. These coordinates form the $(N-1)$-dimensional vector $\boldsymbol{\rho}$. Its length is given by the hyperradius $\rho$,

$$\rho^2 = \sum_{i=1}^{N-1} \xi_i^2 = \sum_{i=1}^{N} r_i^2 - R^2 = \frac{1}{N} \sum_{i<j}^{N} (r_i - r_j)^2, \tag{2}$$

where $R = (\sum_{i=1}^{N} r_i)/\sqrt{N}$ is the normalised coordinate of the centre of mass. The other hyperspherical coordinates, the hyperangles $\hat{\rho} \equiv \{\theta_1, \theta_2, \ldots, \theta_{N-2}\}$, can be chosen in many different ways but they are not used in the present approach.

2.2 Schrödinger Equation in Hyperspherical Harmonics Basis

The kinetic-energy operator $\hat{T}$ in hyperspherical coordinates is split into hyperradial and hyperangular parts,

$$\hat{T} = -\frac{\hbar^2}{2m} \left( \frac{1}{\rho^{n-1}} \frac{\partial}{\partial \rho} \left( \rho^{n-1} \frac{\partial}{\partial \rho} \right) - \frac{1}{\rho^2} \Delta_{\hat{\rho}} \right), \tag{3}$$

where $n = N - 1$ is the dimension of the space formed by the Jacobi coordinates. The eigenfunctions $Y_{K\gamma}(\hat{\rho})$ of the hyperangular part $\Delta_{\hat{\rho}}$,

$$\Delta_{\hat{\rho}} Y_{K\gamma}(\hat{\rho}) = K(K + n - 2) Y_{K\gamma}(\hat{\rho}), \tag{4}$$

where $K$ is the hypermomentum, form a complete orthonormal set. They can be highly degenerate and different harmonics belonging to the same $K$ are labelled by $\gamma$, which represents all the angular momentum quantum numbers. Below, they are constructed symmetrical with respect to permutations of particles in order to be able to describe bosonic systems.

The wave function of the $N$-body system in these coordinates, $\Psi(\boldsymbol{\rho})$, can be expanded over complete sets of hyperspherical harmonics,

$$\Psi(\boldsymbol{\rho}) = \rho^{-(n-1)/2} \sum_{K\gamma} \chi_{K\gamma}(\rho) Y_{K\gamma}(\hat{\rho}). \qquad (5)$$

The hyperradial parts $\chi_{K\gamma}(\rho)$ satisfy the infinite set of coupled differential equations

$$\left(-\frac{d^2}{d\rho^2} + \frac{\mathcal{L}_K(\mathcal{L}_K+1)}{\rho^2} - \frac{2m}{\hbar^2}\left(E - V_{K\gamma,K\gamma}(\rho)\right)\right)\chi_{K\gamma}(\rho)$$
$$= -\frac{2m}{\hbar^2} \sum_{K'\gamma' \neq K\gamma} V_{K\gamma,K'\gamma'}(\rho)\chi_{K'\gamma'}(\rho), \qquad (6)$$

where

$$\mathcal{L}_K = K + \tfrac{1}{2}(N-4) \qquad (7)$$

is the generalized angular momentum, $m$ is the mass of the particle and the hyperradial potentials $V_{K\gamma,K'\gamma'}(\rho)$ are the matrix elements of the interaction potential $\hat{V}$ between HHs,

$$V_{K\gamma,K'\gamma'}(\rho) = \langle Y_{K\gamma}(\hat{\rho}) | \hat{V} | Y_{K'\gamma'}(\hat{\rho}) \rangle. \qquad (8)$$

2.3 Construction of the Hyperspherical Harmonics

Following Refs. [5–7], we construct the hyperspherical harmonics $Y_{K\gamma}(\hat{\rho})$ from linear combinations of eigenfunctions of the harmonic-oscillator Hamiltonian,

$$H = \sum_{i=1}^{N}\left(-\frac{\hbar^2}{2m}\frac{\partial^2}{\partial r_i^2} + \frac{1}{2}m\omega^2 r_i^2\right), \qquad (9)$$

where $\omega$ is the oscillator frequency. This Hamiltonian can be rewritten in hyperspherical $\rho$, $\hat{\rho}$ and centre-of-mass $R$ coordinates as

$$H = -\frac{\hbar^2}{2m}\frac{\partial^2}{\partial R^2} + \frac{1}{2}m\omega^2 R^2 + \hat{T} + \frac{1}{2}m\omega^2\rho^2. \qquad (10)$$

The eigenfunctions of the Hamiltonian (10) corresponding to the lowest energy of the centre-of-mass motion and the internal energy eigenvalues $(2\kappa + K + n/2)\hbar\omega$ are factorised as

$$\Psi^{(0)}_{\kappa K\gamma}(R,\rho,\hat{\rho}) = \Phi_0^{\text{c.m.}}(R)\Phi_{\kappa K}(\rho) Y_{K\gamma}(\hat{\rho}), \qquad (11)$$

where

$$\Phi_0^{\text{c.m.}}(R) = \pi^{-1/4} b^{-1/2} e^{-R^2/2b^2} \qquad (12)$$

is the lowest-energy centre-of-mass wave function and the hyperradial wave function $\Phi_{\kappa K}(\rho)$ is [1]

$$\Phi_{\kappa K}(\rho) = b^{-n/2}\left(\frac{2\kappa!}{\Gamma(\kappa + K + n/2)}\right)^{1/2}\left(\frac{\rho}{b}\right)^K L_\kappa^{K+(n-2)/2}\left(-\frac{\rho^2}{b^2}\right) e^{-\rho^2/2b^2}. \qquad (13)$$

In Eqs. (12) and (13), $b = \sqrt{\hbar/m\omega}$ is the oscillator radius. It follows from (11) that

$$Y_{K\gamma}(\hat{\rho}) = \frac{\Psi^{(0)}_{\kappa K\gamma}(R,\rho,\hat{\rho})}{\Phi_0(R)\Phi_{\kappa K}(\rho)} \qquad (14)$$

**Table 1** Number of channels $\gamma$ for selected values of $K$ for different boson numbers $N$

| $K$ | $N=4$ | $N=5$ | $N=6$ | $N=8$ | $N=10$ | $N=20$ | $N=50$ | $N=100$ |
|---|---|---|---|---|---|---|---|---|
| 10 | 1 | 2 | 3 | 4 | 5 | 5 | 5 | 5 |
| 20 | 2 | 6 | 11 | 24 | 34 | 49 | 49 | 49 |
| 30 | 3 | 11 | 29 | 84 | 153 | 316 | 331 | 331 |
| 40 | 4 | 18 | 54 | 227 | 511 | | | |

for arbitrary $\kappa$ and $\omega$. Following [6], we choose $\kappa = 0$. Since the wave function $\Psi^{(0)}_{0K\gamma}(R, \rho, \hat{\rho})$ is also an eigenfunction of the Hamiltonian (9), it can be represented as

$$\Psi^{(0)}_{0K\gamma}(R, \rho, \hat{\rho}) = \sum_{\mu} C^{\mu}_{K\gamma} \Phi_{K\mu}(r_1, r_2, \ldots, r_n), \tag{15}$$

where $\Phi_{K\mu}(r_1, r_2, \ldots, r_n)$ are symmetrized products of single-particle wave functions $\phi_{m_i}(r_i)$ defined as

$$\phi_m(r) = \frac{\pi^{-1/4}}{\sqrt{2^m m!\,b}} \exp\left(-\frac{r^2}{2b^2}\right) H_m\left(\frac{r}{b}\right), \tag{16}$$

with $H_m$ being the Hermite polynomial of order $m$. In (15), the index $\mu$ represents $\{m_1, m_2, \ldots, m_N\}$ and $m_1 + m_2 + \cdots + m_N = K$. Thus, the hyperspherical harmonics (14) have the following structure:

$$Y_{K\gamma}(\hat{\rho}) = \frac{\sum_{\mu} C^{\mu}_{K\gamma} \Phi_{K\mu}(r_1, r_2, \ldots, r_n)}{\Phi_0(R)\Phi_{0K}(\rho)},$$

where the coefficients $C^{\mu}_{K\gamma}$ are constructed in such a way that only the $0s$ oscillation of the centre-of-mass is present in $\Psi^{(0)}_{\kappa K\gamma}(R, \rho, \hat{\rho})$ and that the HH $Y_{K\gamma}(\hat{\rho})$ is an eigenfunction of $\Delta_{\hat{\rho}}$ (see "Appendices A and B" for more details).

The number of positive-parity hyperspherical harmonics for some values of $K$ is shown in Table 1 for several many-boson systems. For $N \geq 4$, the number of positive-parity HHs is non-zero for any $K$ except $K = 2$. For $N = 3$, only those HHs where $K$ is a multiple of six exist, one HH per $K$. This does not contradict the HH basis from [3], where the HHs are just the functions $e^{iK\theta}$ of the angle $\theta = \arctan(\xi_2/\xi_1)$ with $K$ values of $0, \pm 6, \pm 12, \pm 18, \ldots$. Indeed, transforming the $e^{iK\theta}$ basis into a $\cos K\theta$ and $\sin K\theta$ basis results in non-zero potential matrix elements only for $\cos K\theta$, leaving effectively only one HH channel per $K$, i.e.,

$$Y_K(\hat{\rho}) = [(1 + \delta_{K0})\pi]^{-1/2} \cos K\theta. \tag{17}$$

### 2.4 Hyperradial Potentials

To calculate the hyperradial potentials $V_{K\gamma, K'\gamma'}(\rho)$ that enter Eq. (6), we use the technique developed in Ref. [6] to obtain

$$V_{K\gamma, K'\gamma'}(\rho) = \frac{\sqrt{\Gamma(K+n/2)\Gamma(K'+n/2)}}{\rho^{K+K'+n-2}} \\ \times \frac{1}{2\pi i} \int_{-i\infty}^{i\infty} ds\, e^{s\rho^2} s^{-(K+K'+n)/2} V_{K\gamma, K'\gamma'}(s), \tag{18}$$

in which the integration path bypasses the origin in the counterclockwise direction. The integral in Eq. (18) is the inverse Laplace transform of the function $s^{-(K+K'+n)/2} V_{K\gamma, K'\gamma'}(s)$, many of which are tabulated in Ref. [13]. The potentials

$$V_{K\gamma, K'\gamma'}(s) = \sum_{\mu\mu'} C^{\mu}_{K\gamma} C^{\mu'}_{K'\gamma'} \langle \Phi_{K\mu} | \hat{V} | \Phi_{K'\mu'} \rangle \tag{19}$$

contain the usual matrix elements $\langle\Phi_{K\mu}|\hat{V}|\Phi_{K'\mu'}\rangle$ in the oscillator basis, which are calculated using standard many-body techniques for $b = s^{-1/2}$. The integration over $s$ can be easily performed analytically before making the integration over $\{r_i\}$. For some potentials, among which the contact and Gaussian ones, it is possible to make the integration over $\{r_i\}$ first and then integrate over $s$ analytically using available tables of inverse Laplace transforms (see "Appendix C").

## 3 Solving the Hyperradial Equations

To solve the system (6) of coupled hyperradial equations, we use the Lagrange-mesh method [8,10]. This method is an approximate variational calculation making use of $M$ Lagrange basis functions related to the $M$ mesh points of an associated Gauss quadrature. As the matrix elements are evaluated with this quadrature, the resulting equations take the form of a calculation on a mesh. In spite of this approximation, numerical tests on a variety of problems show that a Lagrange-mesh calculation is essentially as accurate as the corresponding variational calculation involving numerically exact matrix elements [11] (see also Sect. 4.2). This striking property is obtained provided that no singularities appear in the matrix elements and that the behaviours of the wave function at the origin and at infinity are well simulated by the basis. The problem raised by the existence of a singularity can be solved with a regularization technique [9].

3.1 Lagrange-mesh Method

For a variable varying from zero to infinity, it is convenient to use the $M$ regularized Lagrange–Laguerre functions

$$\hat{f}_j(x) = (-1)^j (h_M^\alpha x_j)^{-1/2} \frac{L_M^\alpha(x)}{x - x_j} x^{\alpha/2+1} e^{-x/2}, \quad (20)$$

where $L_M^\alpha(x)$ is a generalized Laguerre polynomial [14] depending on parameter $\alpha > -1$ and $h_M^\alpha = \Gamma(M + \alpha + 1)/M!$ is the square of its norm. The associated zeros $x_j$, $j = 1, \ldots, M$, are given by

$$L_M^\alpha(x_j) = 0. \quad (21)$$

They correspond to the mesh points of the Gauss-Laguerre quadrature

$$\int_0^\infty g(x)dx \approx \sum_{k=1}^M \lambda_k g(x_k), \quad (22)$$

where the $\lambda_k$ are the Gauss weights. The $x_k$ and $\lambda_k$ depend on $\alpha$ and $M$. This quadrature method is exact when $g(x)$ is the product of a polynomial of degree at most $2M - 1$ by the weight function of the Laguerre polynomials

$$w(x) = x^\alpha e^{-x}. \quad (23)$$

The Lagrange functions (20) are thus the product of the square root of the weight function $w(x)$ by $x$ and by a polynomial of degree $M - 1$. They vanish at the origin and verify the Lagrange property

$$\hat{f}_j(x_i) = \lambda_i^{-1/2}\delta_{ij}. \quad (24)$$

Without the factor $x$, the Lagrange functions would be exactly orthonormal. The usefulness of the regularizing factor $x$ is explained below. The functions (20) are not orthogonal but are still orthonormal at the Gauss quadrature (22). As explained in Ref. [10], this basis will be treated as orthonormal in the following without loss of accuracy.

The hyperradial functions are expanded as

$$\chi_{K\gamma}(\rho) = h^{-1/2} \sum_{j=1}^M c_{K\gamma j} \hat{f}_j(\rho/h), \quad (25)$$

where $h$ is a scaling parameter and $h^{-1/2} \hat{f}_j(\rho/h)$ is normed at the Gauss approximation. Their behaviour at the origin is given by

$$\chi_{K\gamma}(\rho) \underset{\rho \to 0}{\sim} \rho^{\mathcal{L}_K + 1}. \tag{26}$$

For $x \to 0$, the Lagrange functions behave as

$$\hat{f}_j(x) \underset{x \to 0}{\sim} x^{\alpha/2 + 1}. \tag{27}$$

We choose

$$\alpha = 2\mathcal{L}_0 = N - 4 \tag{28}$$

to simulate the behaviour (26) for $K = 0$. This choice is valid for $N \geq 4$. The case $N = 3$ must be treated separately (see Sect. 3.3). The other behaviours (26) can be simulated by linear combinations of Lagrange functions if $M$ is large enough.

As explained above, the overlap matrix elements are approximated with the Gauss quadrature (22) as

$$\int_0^\infty \hat{f}_i(\rho/h) \hat{f}_j(\rho/h) d\rho \approx h \delta_{ij} \tag{29}$$

thanks to the Lagrange property (24) and are thus treated as orthonormal. The crucial simplification of the Lagrange-mesh method is that the matrix elements of the potential are approximated as

$$\int_0^\infty \hat{f}_i(\rho/h) V(\rho) \hat{f}_j(\rho/h) d\rho \approx h V(h x_i) \delta_{ij}. \tag{30}$$

The potential matrix is diagonal. Thanks to the regularization by $x$, the matrix elements of $1/x$ and $1/x^2$ are *exact* with the Gauss quadrature. The hyperradial equations take the form of mesh equations,

$$\sum_{j=1}^M \frac{1}{h^2} \left( T_{ij} + \frac{\mathcal{L}_K(\mathcal{L}_K + 1)}{x_j^2} \delta_{ij} \right) c_{K\gamma j} + \frac{2m}{\hbar^2} (V_{K\gamma, K\gamma}(h x_i) - E) c_{K\gamma i}$$
$$= -\frac{2m}{\hbar^2} \sum_{K'\gamma' \neq K\gamma} V_{K\gamma, K'\gamma'}(h x_i) c_{\gamma' K' i}. \tag{31}$$

The kinetic-energy matrix elements of $-d^2/dx^2$ read at the Gauss approximation [10,15]

$$T_{i \neq j} = (-1)^{i-j} \frac{x_i + x_j}{\sqrt{x_i x_j}(x_i - x_j)^2} \tag{32}$$

and

$$T_{ii} = -\frac{1}{12 x_i^2} \left[ x_i^2 - 2(2M + \alpha + 1) x_i + \alpha^2 - 4 \right]. \tag{33}$$

They are simple expressions involving the Laguerre zeros $x_j$.

Although approximations (29) and (30) are rather poor for individual matrix elements, the resulting Lagrange-mesh calculation can be very accurate [10,11]. This striking property is illustrated here by the results in Sects. 4 and 5.

### 3.2 Hybrid Lagrange-mesh method

For large numbers of bosons, the increasing degeneracies of large $K$ values (see Table 1) lead to strongly increasing sizes of the matrices. Computer times may then forbid reaching a sufficient convergence with respect to $K$. To partly compensate this effect, it would be efficient to use decreasing numbers of mesh points when $K$ increases. If there is no significant loss of accuracy, this would reduce the size of the matrix and thus the computer time. However, the simplicity of (30) is then lost in the calculation of some matrix elements connecting different $K$ values since the associated meshes and Gauss quadratures then differ. These matrix elements can nevertheless be accurately calculated in many cases with a third Gauss quadrature [16]. This leads to a hybrid calculation where some matrix elements are calculated like in the previous subsection and a different method of evaluation is used for the other ones.

Such a method opens the way to different strategies. One can keep a unique value for $\alpha$ as before and vary the numbers of mesh points and scale parameters. One can also vary the $\alpha$ values. For example, one can use basis functions with the correct behaviour (26) at the origin for each partial wave by choosing

$$\alpha = 2\mathcal{L}_K = 2K + N - 4. \tag{34}$$

These hybrid calculations may reduce the matrix size at the cost of a more complicated search for optimal numbers of mesh points and scale parameters. As shown below, the best strategy we have found happens to be quite simple.

With respect to the Lagrange-mesh method, a new type of calculation is necessary. For $\alpha \neq \alpha'$, one must calculate the matrix elements between Lagrange functions,

$$\langle \hat{f}_i^\alpha | V(\rho) | \hat{f}_{i'}^{\alpha'} \rangle = \int_0^\infty h^{-1/2} \hat{f}_i^\alpha(\rho/h) V(\rho) h'^{-1/2} \hat{f}_{i'}^{\alpha'}(\rho/h') d\rho, \tag{35}$$

where the superscripts indicate the parameter of the Laguerre polynomial and where the indices $i$ and $i'$ of the mesh points $x_i$ and $x'_{i'}$ vary from 1 to $M$ and 1 to $M'$, respectively. The basis functions also depend on different scale factors $h$ and $h'$. In addition to the potential, the integrand in this matrix element contains a factor $\rho^{(\alpha+\alpha')/2+2}$ multiplied by a polynomial of degree $M+M'-2$ and the exponential $\exp(-\rho/2h - \rho/2h')$.

In the case $V(\rho) \propto 1/\rho$ which will be encountered below, the integral (35) can be exactly computed with a Gauss-Laguerre quadrature with Laguerre parameter

$$\alpha'' = \tfrac{1}{2}(\alpha + \alpha') \tag{36}$$

and scale parameter

$$h'' = \frac{2hh'}{h + h'} \tag{37}$$

provided the number $M''$ of Gauss mesh points $x''_k$ and weights $\lambda''_k$ verifies

$$M'' \geq \tfrac{1}{2}(M + M'). \tag{38}$$

The matrix element of $1/\rho$ is then exactly given by

$$\left\langle \hat{f}_i^\alpha \left| \frac{1}{\rho} \right| \hat{f}_{i'}^{\alpha'} \right\rangle = \frac{h''}{\sqrt{hh'}} \sum_{k=1}^{M''} \lambda''_k \hat{f}_i^\alpha(h'' x''_k / h) \frac{1}{h'' x''_k} \hat{f}_{i'}^{\alpha'}(h'' x''_k / h'). \tag{39}$$

The choice $\alpha'' = (\alpha + \alpha')/2 + 1$ is also possible.

A similar formula should be accurate for more general hyperradial potentials. If these potentials are regular at the origin, the value of $\alpha''$ may be increased by one unit. The value of $M''$ must be large enough to reach a sufficient accuracy.

### 3.3 Special Treatment of $N = 3$

For $N = 3$, one has $\mathcal{L}_0 = -1/2$ and the $K = 0$ 'centrifugal' term becomes attractive. This gives rise to difficulties in most numerical approaches. The previous methods are not valid since $\alpha = -1$ appears. A special treatment of $K = 0$ is necessary. It is made possible by introducing non-regularized Lagrange functions for $K = 0$ in the hybrid method. As shown below, it leads to a fast and accurate computation.

The behaviour of the hyperradial functions at the origin is given by

$$\chi_K(\rho) \underset{\rho \to 0}{\sim} \rho^{K+1/2}. \tag{40}$$

The function $\chi_0$ thus behaves as $\rho^{1/2}$. The $M$ non-regularized Lagrange functions [8,10] are defined as

$$f_j(x) = \frac{x_j}{x} \hat{f}_j(x) = (-1)^j \left(\frac{x_j}{h_M^\alpha}\right)^{1/2} \frac{L_M^\alpha(x)}{x - x_j} x^{\alpha/2} e^{-x/2}. \tag{41}$$

With

$$\alpha = 1, \tag{42}$$

they have the required behaviour at the origin. With this basis, the matrix elements of $1/x^2$ and of $-d^2/dx^2$ do not exist but the matrix elements of the full kinetic operator converge [10,17] and are given for $i \neq j$ by

$$\left\langle f_i \left| -\frac{d^2}{dx^2} - \frac{1}{4x^2} \right| f_j \right\rangle = (-1)^{i-j} \frac{1}{\sqrt{x_i x_j}} \left( \frac{M+1}{2} - \frac{1}{x_i} - \frac{1}{x_j} + \frac{x_i + x_j}{(x_i - x_j)^2} \right) \tag{43}$$

and for $i = j$ by

$$\left\langle f_i \left| -\frac{d^2}{dx^2} - \frac{1}{4x^2} \right| f_i \right\rangle = -\frac{1}{12x_i^2} [x_i^2 - 10(M+1)x_i + 24]. \tag{44}$$

These expressions replace the parenthesis in the first term of (31) for $K = 0$.

The Gauss quadrature is not accurate for matrix elements of $1/x$ but one can use the exact expression [10,17],

$$\left\langle f_i \left| \frac{1}{x} \right| f_j \right\rangle = \frac{1}{x_i} \delta_{ij} + (-1)^{i-j} \frac{1}{\sqrt{x_i x_j}}. \tag{45}$$

This expression is useful in the case of contact interactions described below. In this case, the potential terms calculated with the Gauss quadrature in (31) must be replaced by exact expressions with the help of (45).

Other $K$ values are treated like in the hybrid method. The expression of non-diagonal matrix elements between $K = 0$ and $K' \neq 0$ must be adapted from (39) by replacing $\hat{f}_i^\alpha$ by $f_i$ for $K = 0$.

Finally, when $V(\rho) \propto 1/\rho$, an exact variational calculation is also possible. All matrix elements for $K = 0$ and all matrix elements of the potential are exact. The overlaps of the regularized functions $\hat{f}_j(x)$ are given by Eq. (3.71) of Ref. [10] and the exact kinetic-energy matrix elements are given by Eq. (3.77). The variational results are then obtained from a generalized eigenvalue problem.

## 4 HH Expansion for Zero-Range Forces

For contact (zero-range) two-body forces ($V_0 > 0$),

$$\hat{V} = -V_0 \sum_{i>j} \delta(r_i - r_j), \tag{46}$$

the exact energy is known analytically [18]. This allows precise tests of the convergence of the HH expansion.

The Schrödinger equation verifies the scaling property

$$V_0 \to \nu V_0, \quad r_i \to r_i/\nu, \quad E \to E/\nu^2. \tag{47}$$

The same property is verified by the system of coupled equations (6) and by its approximation (31), even when the system is truncated at any value of $K$.

The matrix elements of the potential $\langle \Phi_{K\mu} | \hat{V} | \Phi_{K'\mu'} \rangle$ in the oscillator basis are proportional to $\sqrt{s}$, which gives with (19) the hyperradial potentials as

$$V_{K\gamma, K'\gamma'}(\rho) = -\frac{c_{K\gamma, K'\gamma'}}{\rho}. \tag{48}$$

The system (6) looks like a coupled Coulomb problem for a two-body system in three dimensions. The coefficients for the three-boson case in one dimension can be derived from Eq. (3.3) of Ref. [3]. In the basis (17) constructed in the present paper, the non-vanishing coefficients are

$$c_{K,K'} = \frac{3\sqrt{2} V_0 (-1)^{(K+K')/2}}{\pi [(1 + \delta_{K0})(1 + \delta_{K'0})]^{1/2}}, \tag{49}$$

where $K$ and $K'$ are multiples of 6. For all other numbers $N$ of bosons, the $c_{K\gamma, K'\gamma'}$ are obtained from numerical calculations.

### 4.1 $K = 0$ Limit

For $K = 0$, (19) reads with (16),

$$V_{00}(s) = -\frac{N(N-1)}{2} V_0 \int_{-\infty}^{+\infty} [\phi_0(x)]^4 \, dx = -\frac{N(N-1)}{2} V_0 \sqrt{\frac{s}{2\pi}}. \tag{50}$$

With an inverse Laplace transform [14], the corresponding matrix element (18) reads

$$V_{00}(\rho) = -\frac{N(N-1)}{2} \frac{\Gamma(\frac{N-1}{2})}{\Gamma(\frac{N-2}{2})} \frac{V_0}{\sqrt{2\pi}\,\rho}. \tag{51}$$

If we keep in expansion (5) the $K = 0$ term only, then the hyperradial equation (6) is exactly the same as the one for a hydrogen-like atom with non-zero angular momentum $\mathcal{L}_0 = \frac{1}{2}(N - 4)$,

$$\left[ -\frac{d^2}{d\rho^2} + \frac{\mathcal{L}_0(\mathcal{L}_0 + 1)}{\rho^2} - \frac{2m}{\hbar^2} \left( \frac{c_{00}}{\rho} + E_{K=0} \right) \right] \chi_0(\rho) = 0, \tag{52}$$

with

$$c_{00} = \frac{N(N-1)}{2} \frac{\Gamma(\frac{N-1}{2})}{\Gamma(\frac{N-2}{2})} \frac{V_0}{\sqrt{2\pi}}. \tag{53}$$

The solutions of this hydrogen-like equation are well known. Its lowest-energy eigenvalue for each $\mathcal{L}_0$ value is

$$E_{K=0} = -\frac{m}{2\hbar^2} \left( \frac{c_{00}}{\mathcal{L}_0 + 1} \right)^2 = -\frac{2m}{\hbar^2} \left( \frac{c_{00}}{N - 2} \right)^2 \tag{54}$$

and the corresponding wave function is

$$\chi_0(\rho) = \sqrt{\frac{2^{N-1} \lambda}{\Gamma(N-2)}} \, (\lambda \rho)^{\frac{N-2}{2}} e^{-\lambda \rho}, \tag{55}$$

where the inverse 'Bohr radius' reads

$$\lambda = \frac{2m}{\hbar^2} \frac{c_{00}}{N - 2}. \tag{56}$$

**Table 2** Exact energy for an $N$-boson system with contact interactions in comparison with the one obtained in the $K=0$ approximation and with the minimum expectation energy $E_{\text{HO}}^{\min}$ in the harmonic oscillator basis

| $N$ | $E_{\text{exact}}$ | $E_{K=0}$ | $E_{\text{HO}}^{\min}$ |
|---|---|---|---|
| 3 | −0.5 | −0.45595 | −0.35810 |
| 4 | −1.25 | −1.1250 | −0.95493 |
| 5 | −2.5 | −2.2516 | −1.9894 |
| 6 | −4.375 | −3.9551 | −3.5810 |
| 7 | −7.0 | −6.3549 | −5.8489 |
| 8 | −10.5 | −9.5703 | −8.9127 |
| 9 | −15.0 | −13.721 | −12.892 |
| 10 | −20.625 | −18.926 | −17.905 |
| 20 | −166.25 | −155.23 | −151.20 |
| 100 | −20831.25 | −19795 | −19695 |

All energies are shown as ratios to $2mV_0^2/\hbar^2$

The exact analytic solution for a zero-range force is known as [18]

$$E_{\text{exact}} = -\frac{2m}{\hbar^2}\frac{V_0^2 N(N^2-1)}{48}. \tag{57}$$

In the $N \to \infty$ limit, one has

$$E_{K=0} \to \frac{48}{16\pi} E_{\text{exact}} \to -\frac{2m}{\hbar^2}\frac{V_0^2 N^3}{16\pi}.$$

So, both $E_{K=0}$ and $E_{\text{exact}}$ have the same cubic asymptotic behaviour. The $N^2$ behaviour of $E_{K=0}$ in Ref. [2] comes from a missing factor $\Gamma(\frac{N-1}{2})/\Gamma(\frac{N-2}{2})$ due to the absence of $\int d\Omega_{N-2}/\int d\Omega_{N-1}$, where $\Omega_i$ is the solid angle in an $(i+1)$-dimensional space, in their analytical expression for the $K=0$ energy. Indeed, for $N$ large, one has

$$\left(\frac{\Gamma(\frac{N-1}{2})}{\Gamma(\frac{N-2}{2})}\right)^2 \approx \frac{\Gamma(\frac{N-1}{2})}{\Gamma(\frac{N-2}{2})}\frac{\Gamma(\frac{N}{2})}{\Gamma(\frac{N-1}{2})} \approx \frac{N}{2}.$$

The evaluation of the solid angle $\int d\Omega_{N-1}$ is explained in "Appendix D".

Interestingly, the same $N^3$ behaviour can be obtained by minimizing the expectation value of the two-body interaction in the lowest-order oscillator basis. It gives

$$E_{\text{HO}}^{\min} = -\frac{2m}{\hbar^2}\frac{V_0^2 N^2(N-1)}{16\pi} \to E_{K=0},$$

from which it follows that $E_{\text{HO}}^{\min}$ is always higher than $E_{K=0}$ but approaches the latter asymptotically. Both $E_{\text{HO}}^{\min}$ and $E_{K=0}$ are shown in Table 2 in comparison with the exact solution $E_{\text{exact}}$.

### 4.2 Convergence of HH Expansion for Contact Interaction in a Three-Boson Case

The convergence of the HH expansion for a three-body system in one dimension with the contact interaction was shown to be slow [3]. It was suggested there that this convergence has an exponential behaviour. We confirm the slow convergence but we disagree with its exponential form. Because of the small size of the matrices involved in the 3-body case and availability of exact expressions for the hyperradial potentials, it is possible to reach very high values of $K_{\max}$, such as $K_{\max}=6000$, so that convergence can be studied precisely.

Because of the scaling property (47), the convergence of the HH expansion for a contact interaction is independent of the choice of $V_0$ for any fixed $N$. Therefore, we choose the parameters

$$\frac{\hbar^2}{2m}=1, \quad V_0=1. \tag{58}$$

**Table 3** Convergence of energies of hybrid Lagrange-mesh and variational calculations for $N = 3$ bosons at $K_{\text{max}} = 120$ and 1200 with respect to the number of mesh points $M = M'$

| $M = M'$ | $K_{\text{max}} = 120$ | | $K_{\text{max}} = 1200$ | |
|---|---|---|---|---|
| | Hybrid | Variational | Hybrid | Variational |
| 2 | −0.4985108990 | −0.4968799048 | −0.4998465087 | −0.4981845707 |
| 3 | −0.4985114348 | −0.4985090735 | −0.4998472894 | −0.4998444821 |
| 4 | −0.4985114348 | −0.4985114307 | −0.4998472898 | −0.4998472819 |
| 5 | −0.4985114348 | −0.4985114348 | −0.4998472898 | −0.4998472898 |

The scale factors are $h = h' = 0.74$. For information, the exact value is $-0.5$

As shown for hydrogen-like atoms in Ref. [10], a Lagrange-mesh or variational calculation with a single Lagrange function can give the exact lowest energy for $K = 0$ since the corresponding wave function can be simulated by this Lagrange function (compare (55) with (20) or (41) for $M = 1$) and the Gauss quadrature is exact. This is obtained with the scale parameter

$$h = \frac{1}{2\lambda} = (N - 2)\frac{\hbar^2}{4mc_{00}}. \tag{59}$$

This should thus be a good scale parameter for the full calculation.

In calculating $E(K_{\text{max}})$, we had to use a modification of the Lagrange-mesh approach because its original version is not valid due to the attractive 'centrifugal' term at $K = 0$. This case is treated as explained in Sect. 3.3. We have tested different strategies and found a simple and efficient one. Different bases are used, non regularized for $K = 0$ (Eq. (41) with $\alpha = 1$) and regularized for $K' \neq 0$ (Eq. (20) with $\alpha' = 1$). The numbers of basis functions are $M$ and $M'$ and the scale parameters are $h = h' = 0.74$, rounded from (59). All matrix elements involving $K = 0$ are calculated exactly, as well as all integrals of type (39). The size of the matrix is $M' \times K_{\text{max}}/6 + M$.

The convergence with respect to the number of mesh points is fast as shown in Table 3. An accuracy of $10^{-10}$ is reached with $M = M' = 4$. The validity of this hybrid mesh calculation can be tested. Indeed, one can easily perform an exact variational calculation by exactly evaluating the remaining matrix elements, i.e., the overlaps and the kinetic-energy matrix elements of regularized functions. This leads to a generalized eigenvalue problem. The comparison in Table 3 shows that the hybrid Lagrange-mesh results converge faster than the variational results but that the difference is small and decreases with increasing $M = M'$. The slower convergence of the variational method is probably be due to the non-orthogonality of the basis.

We have found that the behaviour of $E(K_{\text{max}})$ is well described by

$$E(K_{\text{max}}) = E(\infty) + \frac{a_0}{a_1 + K_{\text{max}}}, \tag{60}$$

where $E(\infty) \equiv E_{\text{exact}}$, with $a_0 = 0.183786$ and $a_1 = 3.45912$. This is illustrated in Fig. 1, which shows that the ratio $1/(E(K_{\text{max}}) - E(\infty))$ is practically a linear function over the huge interval $30 \leq K_{\text{max}} \leq 6000$. This ratio should have increased exponentially if the convergence was exponential.

### 4.3 Convergence of HH Expansion for Contact Interaction for $N > 3$

Here we consider the typical values $N = 5, 20$ and $100$ for which we discuss the convergence with respect to (i) the choice of the number of mesh points at a fixed $K_{\text{max}}$ and (ii) the choice of $K_{\text{max}}$ for an optimal number of Lagrange-mesh points.

Very accurate results are obtained with the simple Lagrange-mesh method of Sect. 3.1 for all $K$ values with small number of mesh points and a value of the scale parameter deduced from (59). This value can be rounded without loss of accuracy: $h = 0.33$ for $N = 5$, $h = 0.04$ for $N = 20$ and $h = 0.0035$ for $N = 100$. The convergence with respect to various strategies is analysed in Table 4 at $K_{\text{max}} = 20$ for the three cases. With only two mesh points per partial wave, the relative error is of the order of $10^{-4}$. This accuracy is obtained because of the exactness of the $K = 0$ contribution. With $M = 3$, the error drops to about $10^{-8}$ for $N = 5$, $10^{-6}$ for $N = 20$ and $10^{-5}$ for $N = 100$. All displayed digits are stable for $N = 5$ and 20 with $M = 4$ mesh

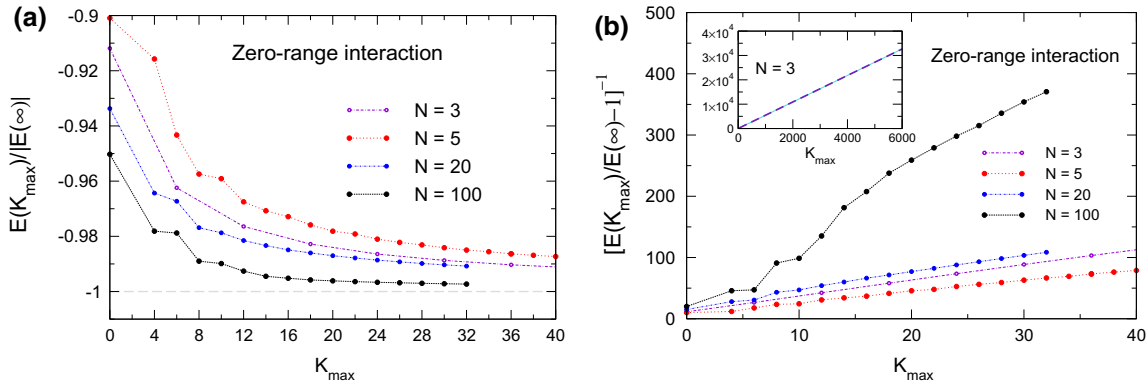

**Fig. 1 a** Ratio of energies $E(K_{max})$, obtained for contact interactions in a model space defined by $K_{max}$, to exact solutions $E(\infty)$, given by Eq. (26), and **b** the quantity $1/(E(K_{max}) - E(\infty))$ calculated for $N = 3, 5, 20$ and $100$. The inset in **b** shows the same value for $N = 3$ at a larger interval of $K_{max}$

**Table 4** Convergence of energies of Lagrange-mesh calculations with respect to $M$ for $N = 5, 20$ and $100$ at $K_{max} = 20$

| $M$ | $N = 5$ | $N = 20$ | $N = 100$ |
|---|---|---|---|
| 2 | −2.44446257 | −164.084958 | −20747.6980 |
| 3 | −2.44463789 | −164.090522 | −20750.6654 |
| 4 | −2.44463792 | −164.090542 | −20750.8121 |
| 5 | −2.44463792 | −164.090542 | −20750.8165 |
| H1 | −2.44463727 | −164.090504 | −20750.7011 |
| H2 | −2.44463792 | −164.090539 | −20750.8135 |
| Exact | −2.5 | −166.25 | −20831.25 |

The respective scale factors are $h = 0.33, 0.04$ and $0.0035$. Comparison with hybrid calculations: H1 where $M = 4$ from $K = 0$ to 10 and $M = 3$ beyond and H2 where, in addition, the scale factor as been increased for $K > 10$ as $h = 0.34, 0.041$ and $0.0038$, respectively. The exact values are given for information points and for $N = 100$ with $M = 5$ mesh points. In any case, the errors for $M \geq 3$ are negligible with respect to the errors due to the $K$ truncation.

The size of the matrix strongly increases with $M$ as illustrated for some $K$ values by examples of degeneracies in Table 1. In the following, we shall truncate the case with $N = 5$ bosons at $K_{max} = 40$. The sum of the degeneracies over $K$ is 145 and the matrix size is thus $M \times 145$. The matrix sizes become much larger when the number of bosons increases. For $N = 20$ at $K_{max} = 32$, the matrix size is $M \times 1451$. For $N = 100$ at the same $K_{max}$, the matrix size is $M \times 1507$.

A hybrid treatment reducing these sizes would thus be helpful. We have compared various strategies and found that the choice (34) leads to complicated searches for optimal sets of $h$ values, without significant improvement. The simplest hybrid strategy is to reduce the number of mesh points beyond some $K$ value. The reduction from $M = 4$ to $M = 3$ beyond $K = 10$ is tested as H1 in Table 4 for $K_{max} = 20$. There is a little but not significant loss of accuracy. It can be partly compensated by a slight increase of $h$ as shown in row H2. For high $K_{max}$ values, this can reduce the diagonalisation time in a useful way. For $N = 20$, the size is reduced from 5804 to 4365 with a negligible loss of accuracy. For $N = 100$, the matrix size is reduced from 6028 to 4533.

The ratio $E(K_{max})/E(\infty)$ is shown in Fig. 1 as a function of the model-space $K_{max}$ for $N = 5, 20$ and 100. The $K_{max}$ value is 40 for $N = 5$ and 32 for $N = 20$ and 100. Figure 1 shows a very slow convergence for all selected $N$. However, for larger $N$, the energies for a fixed $E(K_{max})$ are closer to $E(\infty)$. We have also plotted the $1/(E(K_{max}) - E(\infty))$ values. Similar to the three-body case, their behaviour is very close to linear at large $K_{max}$. However, a close look shows small-scale irregularities that prevent to unambiguously determine the parameters of such linear functions. If $E(K_{max})$ is indeed governed by Eq. (60) then its convergence within five significant digits would be achievable in a huge model space with $K_{max}$ as large as several hundreds. Working with such model spaces is not realistic for $N \geq 4$.

We tried to parametrize $E(K_{max})$ at $K_{max} \geq 20$ by other functional forms, exponential and power functions. Exponential extrapolations result in asymptotic values of $E$ which are slightly higher (of the order of 0.2%) than the exact solutions. The power extrapolations give better asymptotic values of the energies. However,

**Table 5** Converged and extrapolated energies $E(\infty)$ (in K) for an $N$-boson system with a Gaussian interaction of range $a$ (in a.u.) and depth $V_g = V_0/(\sqrt{\pi}a)$ that gives the same volume integral as a zero-range force of strength $V_0$

| $N$ | $a = 0$ | $a = 0.05$ | $a = 0.1$ | $a = 0.2$ | $a = 0.5$ | $a = 1$ |
|---|---|---|---|---|---|---|
| 3 | −2.3105 | −2.293 | −2.260 | −2.223 | −2.1064 | −1.9325 |
| 5 | −11.552 | −11.4<u>2</u> | −11.3<u>1</u> | −11.08<u>7</u> | −10.386 | −9.2852 |
| 20 | −768.23 | −75<u>7</u> | −739.<u>9</u> | −695.88 | −562.71 | −417.70 |
| 100 | −96260 | −8643<u>4</u> | −72883 | −54292 | −31070 | −18552 |

The calculated values are shown for $V_0 = 10$ K. The underlined digits are uncertain by 1–2 units their accurate determination is problematic within the model space considered because of the small-scale irregularities.

## 5 HH expansion for Gaussian Two-Body Interaction

For a Gaussian two-body interaction with range $a$ and depth $V_g$,

$$\hat{V} = -V_g \sum_{i>j} e^{-(r_i - r_j)^2/a^2}, \tag{61}$$

the convergence of the HH expansion is determined by $V_g a^2$, because of the scaling property $r_i \to ar_i$, $V_g \to V_g a^2$, $E \to Ea^2$. The energies $E(K_{\max})$ converge faster with larger $V_g a^2$ and larger $N$, which is exemplified below in Fig. 2. To compare this convergence with that for the contact interaction, $V_g$ has been chosen as $V_0/(\sqrt{\pi}a)$ so that both the contact and Gaussian interactions have the same volume integral.

For $N = 3$, the non-vanishing hyperradial potentials are obtained by a direct calculation of (8) using basis (17) as

$$V_{K,K'}(\rho) = -\frac{3V_g(-1)^{(K+K')/2}}{[(1+\delta_{K0})(1+\delta_{K'0})]^{1/2}}$$
$$\times e^{-\rho^2/a^2}[I_{|K-K'|/2}(\rho^2/a^2) + I_{(K+K')/2}(\rho^2/a^2)], \tag{62}$$

where $K$ and $K'$ are multiples of 6 and $I_n$ is a modified Bessel function of the first kind [14]. The $K = 0$ kinetic energy is calculated with (43) and (44). For $a \to 0$, one recovers (48) and (49).

As for the contact case ($a = 0$), the Lagrange-mesh approximation (30) is not valid for $a < 1$. All potential matrix elements involving $K = 0$ are thus computed numerically with (39) and $M'' = 40 - 60$. The energies obtained for several two-body interactions that have the same volume integral, but different ranges $a$ are presented in Table 5. The ranges $a$ are given in a.u. while the depth $V_0 = 10$ is given in K and $\hbar^2/m = 43.281307$ is chosen to be the same as in Ref. [4]. The energy for $a = 0$ is exact. With $K_{\max} = 120$, convergence is reached for the displayed digits for the other $a$ values. These results are obtained with $h = 1.6$ given by (59) and $M = 20$ or 30. The binding energy progressively decreases when $a$ increases. The analysis of the $E(K_{\max})$ behaviour suggests that it is determined by the $E(\infty) + a_1/(a_2 + K_{\max})^{a_3}$ law. The coefficients $a_3$ are very sensitive to the number of energies $E(K_{\max})$ used to make an extrapolation but, generally, they increase with the range $a$.

For $N \geq 4$, the Lagrange-mesh method is used. The potential matrix elements are calculated as explained in "Appendix C". For the larger $a$ values ($a \geq 0.5$ a.u.), converged energies are shown. For $a < 0.5$ a.u., extrapolated values are given. The underlined digits in Table 5 are uncertain by 1–2 units due to the uncertainties of the extrapolation. For $N = 5$, the results are obtained with $M = 20$ or 30 and $h = 0.08$. For $N = 20$, $M = 6$ mesh points are enough while the optimal $h$ increases progressively from 0.08 to 0.12. For $N = 100$, $M = 6$ mesh points are used for $a \leq 0.2$, $M = 8$ for $a = 0.5$ and $M = 10$ for $a = 1$ while $h$ increases from 0.009 to 0.025. One can see that, with increasing $N$, the energy of the $N$-boson system deviates more strongly from that obtained with the contact interaction as the interaction range gets larger. At $a = 1$, $K_{\max} = 6$ becomes a good approximation with an accuracy better than $10^{-4}$ for $N \geq 20$. For $a > 3$, the accuracy of the $K = 0$ approximation is better than $10^{-3}$.

The $E(K_{\max})$ values show more irregularities with increasing $N$ so that it was not possible to determine an exact functional form of the $E(K_{\max})$ behaviour. The inverse power law, $E(\infty) + a_1/(K_{\max})^{a_2}$, however,

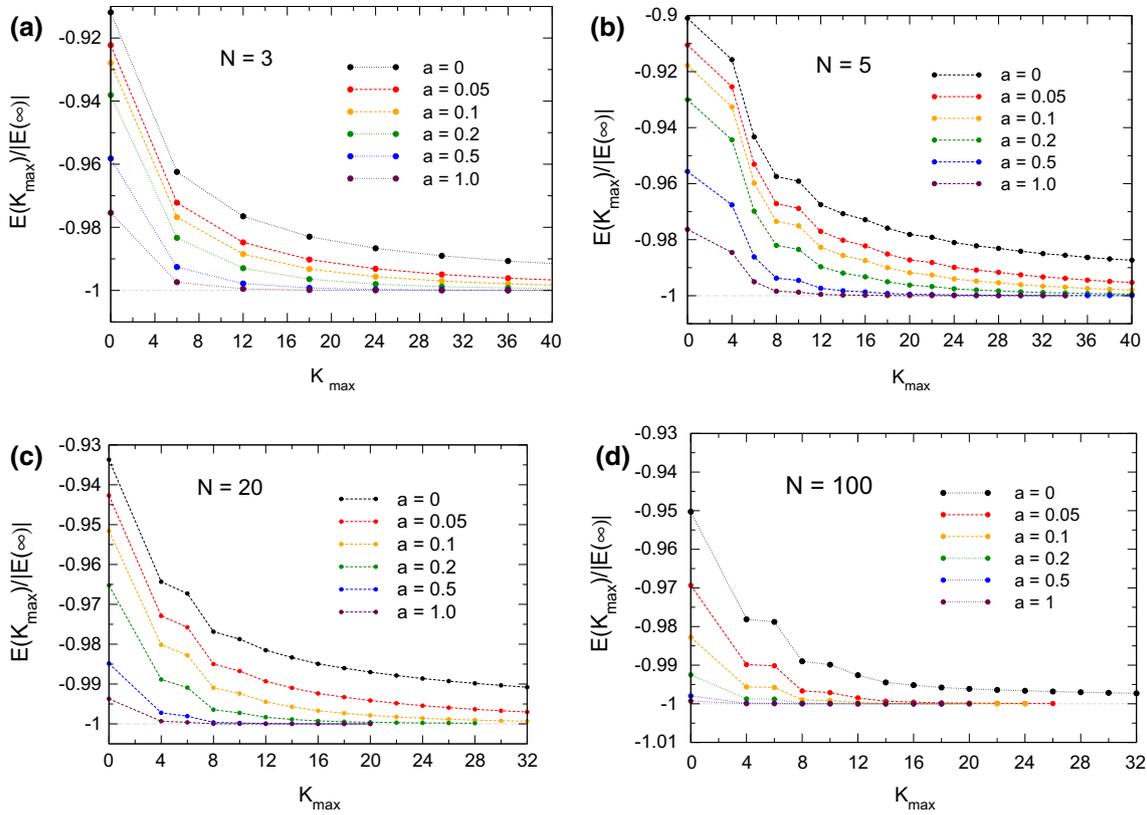

**Fig. 2** Ratio of energies $E(K_{\max})$, obtained for Gaussian interactions of range $a$ in a model space defined by $K_{\max}$, to extrapolated solutions $E(\infty)$ from Table 2 for $N = 3, 5, 20$ and $100$. For comparison, calculations with the corresponding zero-range interaction are also presented

seems to be appropriate to describe $E(K_{\max})$ at $20 \leq K_{\max} \leq 40$, and it was used to determine the $E(\infty)$ values shown in Table 5. To illustrate the $E(K_{\max})$ behaviour, the ratios of energies $E(K_{\max})$ to the extrapolated values $E$ from Table 5 are displayed as a function of $K_{\max}$ in Fig. 2. The accuracy of these values depends on $V_g a^2$ but in all cases their uncertainties are too small to be seen.

## 6 Summary

The hyperspherical-harmonics expansion for bosonic systems in one dimension is revisited and formulated in terms of an expansion on a single-particle oscillator basis. The hyperspherical harmonics are expressed via linear combinations of symmetrized products of single-particle states that contain only the lowest-energy centre-of-mass motion and do not contain hyperradial excitations. They are constructed by diagonalising the matrices of the square of the centre-of-mass coordinate and the hyperangular part of the multidimensional Laplacian. Then the hyperradial potentials are calculated with known methods of many-body physics that use an oscillator basis.

The HH expansion reduces the many-body problem into a one-dimensional coupled-channel problem. We have solved this problem using the Lagrange-mesh method, often with surprisingly small bases. Its convergence has been tested by comparison with the exact analytical ground-state energy for a contact interaction. We found that, in some cases, an extension of the method mixing different meshes can reduce computational times. The $N = 3$ case is treated separately with an exact evaluation of the $K = 0$ component. It easily allows a comparison with an exact variational calculation using the same Lagrange basis. Both methods converge very fast with the number of basis functions and agree.

The HH expansion was applied to treat several bosonic systems interacting with contact and Gaussian forces. It is shown that, for contact interactions, the lowest-order energy in the HH expansion has the correct $N^3$-dependence at large $N$ and is asymptotically higher than the known exact solution by a factor $48/16\pi \approx 0.955$.

The convergence on $K_{\max}$ for these interactions is slow and can be described by an $E(\infty) + a_0/(a_1 + K_{\max})$ dependence. This implies that, to achieve numerically converged values by the HH expansion, very large $K$ values, of an order of a few hundreds, would be needed.

Numerical studies of the HH expansion, carried out for three, five, twenty and hundred bosons with a Gaussian interaction, have shown that their HH convergence rate is strongly determined by the Gaussian range. For a sufficiently large range, the $K_{\max} = 0$ energy can already be close to the converged value. For smaller ranges the convergence deteriorates and extrapolations are needed to get a converged value. The functional behaviour of $E$ with respect to $K_{\max}$ has been found to be reasonably described by an inverse power law at $20 \leq K_{\max} \leq 40$, which makes it very difficult to obtain more than a few converged digits by further basis expansion, except for $N = 3$. As in the three-dimensional cases from [4,5], the convergence gets better with larger boson numbers $N$.

**Appendix**

A Removing the Centre-of-Mass

To guarantee that centre-of mass excitations are absent in the wave function $\Psi^{(0)}_{\kappa K \gamma}(R, \rho, \hat{\rho})$, the expansion coefficients $C^\mu_{K\gamma}$ in (15) are chosen as the eigenvectors of the matrix $R^2$ of the centre-of-mass coordinate squared corresponding to the wave function (12) of the lowest eigenvalue with $s = 1/b^2$,

$$\langle \Phi^{\text{c.m.}}_0 | R^2 | \Phi^{\text{c.m.}}_0 \rangle = \frac{1}{2s}.$$

The operator $R^2$ in individual particle coordinates reads

$$R^2 = \frac{1}{N}\left(\sum_{i=1}^{N} r_i^2 + 2\sum_{i<j}^{N} r_i r_j\right).$$

The calculation of its matrix elements in the oscillator basis involves one-body matrix elements of $r_i^2$,

$$\begin{aligned}
\langle \phi_m | r^2 | \phi_{m'} \rangle &= \frac{1}{2s}(2m + 1), & m' &= m, \\
&= \frac{1}{2s}\sqrt{m(m-1)}, & m' &= m - 2, \\
&= \frac{1}{2s}\sqrt{(m+1)(m+2)}, & m' &= m + 2,
\end{aligned} \quad (A.1)$$

and two-body matrix elements of $r_i r_j$ which include matrix elements of $r$,

$$\langle \phi_m | r | \phi_{m'} \rangle = \sqrt{\frac{1}{2s}}(\sqrt{m'+1}\delta_{m,m'+1} + \sqrt{m'}\delta_{m,m'-1}). \quad (A.2)$$

B Constructing States with Well-Defined Hypermomentum

According to Ref. [19] the angular part of the kinetic energy operator associated with $N$ arbitrary coordinates $\{x_1, x_2, \ldots, x_N\}$ is

$$\Delta_{\Omega_N} = -\sum_{i<j}^{N}\left(x_i \frac{\partial}{\partial x_j} - x_j \frac{\partial}{\partial x_i}\right)^2. \quad (B.1)$$

If the first $N - 1$ of these coordinates are the Jacobi coordinates (1) and the last one is the coordinate of the centre-of-mass, $x_N = R$, then Eq. (B.1) can be split into two parts,

$$\Delta_{\Omega_N} = \Delta_{\Omega_{N-1}} + \Lambda_R, \quad (B.2)$$

where the first term is the angular part of the kinetic energy operator associated with Jacobi coordinates, which is exactly the operator $\Delta_{\hat{\rho}}$,

$$\Delta_{\hat{\rho}} \equiv \Delta_{\Omega_{N-1}} = -\sum_{i<j}^{N-1}\left(\xi_i \frac{\partial}{\partial \xi_j} - \xi_j \frac{\partial}{\partial \xi_i}\right)^2, \tag{B.3}$$

and the second one depends on the centre-of-mass coordinate,

$$\Lambda_R = -\sum_{i=1}^{N-1}\left(\xi_i \frac{\partial}{\partial R} - R \frac{\partial}{\partial \xi_i}\right)^2. \tag{B.4}$$

In order the wave function $\Psi^{(0)}_{\kappa K \gamma}(R, \rho, \hat{\rho})$ to be the eigenfunction of the operator $\Delta_{\hat{\rho}}$, the expansion coefficients $C^{\mu}_{K\gamma}$ in (15) should be chosen as the eigenvectors of the matrix of this operator, calculated in the oscillator basis $\Phi_{K\mu}(r_1, .r_2, \ldots, r_N)$, corresponding to the eigenvalue $K(K + n - 2)$. However, it is inconvenient to calculate matrix elements of the operator defined in Jacobi coordinates by (B.3) in a basis of states constructed in individual coordinates. Instead, we calculate $\langle \Phi_{K\mu'} | \Delta_{\Omega_{N-1}} | \Phi_{K\mu} \rangle$ as

$$\langle \Phi_{K\mu'} | \Delta_{\Omega_{N-1}} | \Phi_{K\mu} \rangle = \langle \Phi_{K\mu'} | \Delta_{\Omega_N} | \Phi_{K\mu} \rangle - \langle \Phi_{K\mu'} | \Lambda_R | \Phi_{K\mu} \rangle \tag{B.5}$$

taking advantage that $\Delta_{\Omega_N}$ is represented by $N$ individual coordinates and that the diagonalization of the operator $\Delta_{\Omega_{N-1}}$ takes place after diagonalization of the matrix $R^2$ so that the new basis functions $\Phi^0_{K\nu} = \sum_{\mu} C^{(0)\nu}_{K\mu} \Phi_{K\mu}$ do not contain centre-of-mass excitations. In this case, one has

$$\langle \Phi^0_{K\nu'} | \Delta_{\hat{\rho}} | \Phi^0_{K\nu} \rangle = \langle \Phi^0_{K\nu'} | \Delta_{\Omega_N} | \Phi^0_{K\nu} \rangle - K \delta_{\nu'\nu}. \tag{B.6}$$

The operator $\Delta_{\Omega_N}$ can be rewritten as

$$\Delta_{\Omega_N} = -\sum_{i<j}^{N}\left(r_i^2 \frac{\partial^2}{\partial r_j^2} + r_j^2 \frac{\partial^2}{\partial r_i^2} - 2r_i \frac{\partial}{\partial r_i} r_j \frac{\partial}{\partial r_j}\right) + (N-1)\sum_{i=1}^{N} r_i \frac{\partial}{\partial r_i}. \tag{B.7}$$

To calculate $\langle \Phi_{K\mu'} | \Delta_{\Omega_N} | \Phi_{K\mu} \rangle$, one needs matrix elements of $r^2$ given by Eq. (A.1) and the matrix elements

$$\begin{aligned}\langle \phi_m | \frac{\partial^2}{\partial r^2} | \phi_{m'} \rangle &= -\frac{s}{2}(2m+1), & m' &= m, \\ &= \frac{s}{2}\sqrt{m(m-1)}, & m' &= m-2, \\ &= \frac{s}{2}\sqrt{(m+1)(m+2)}, & m' &= m+2,\end{aligned} \tag{B.8}$$

and

$$\begin{aligned}\langle \phi_m | r \frac{\partial}{\partial r} | \phi_{m'} \rangle &= -\frac{1}{2}, & m' &= m, \\ &= -\frac{1}{2}\sqrt{m(m-1)}, & m' &= m-2, \\ &= \frac{1}{2}\sqrt{(m+1)(m+2)}, & m' &= m+2.\end{aligned} \tag{B.9}$$

## C Matrix Elements for Hyperradial Potentials

In the approach of the present paper, the hyperradial potentials $V_{K\gamma,K'\gamma'}(\rho)$ are obtained by Laplace-transforming the matrix elements of the two-body potentials in an oscillator basis with oscillator radius $b$. The latter are expressed via two-body matrix elements $\langle m_1 m_2 | \hat{V} | m_3 m_4 \rangle$,

$$\langle \Phi_{K\mu} | \hat{V} | \Phi_{K'\mu'} \rangle = \sum_{m_1+m_2+m_3+m_4 \leq K+K'} c^{\mu\mu'}_{m_1 m_2 m_3 m_4} \langle m_1 m_2 | \hat{V} | m_3 m_4 \rangle, \tag{C.1}$$

where $|m\rangle$ stands for a short-hand notation of $\varphi_m(r)$. Using the Talmi transformation

$$\varphi_{m_1}(r_1)\varphi_{m_2}(r_2) = \sum_{mM} \langle Mm | m_1 m_2 \rangle \varphi_M(R) \varphi_m(r), \tag{C.2}$$

where $r = (r_1 - r_2)/\sqrt{2}$, $R = (r_1 + r_2)/\sqrt{2}$ and the coefficients $\langle Mm | m_1 m_2 \rangle$ have the explicit expression [20]

$$\langle Mm | m_1 m_2 \rangle = 2^{-\frac{m_1+m_2}{2}} \sqrt{\frac{m!M!}{m_1!m_2!}} \sum_{i_1+i_2=m} \frac{(-)^{i_2} m_1! m_2!}{i_1!(m_1-i_1)! i_2!(m_2-i_2)!}, \tag{C.3}$$

the two-body matrix element $\langle m_1 m_2 | \hat{V} | m_3 m_4 \rangle$ becomes

$$\langle m_1 m_2 | \hat{V} | m_3 m_4 \rangle = \sum_{mm'MM'\nu\nu'} \langle Mm | m_1 m_3 \rangle \langle M'm' | m_2 m_4 \rangle \langle \nu\nu' | MM' \rangle$$
$$\times \frac{c_\nu}{2} \varphi_m(0) \varphi_{m'}(0) \int_{-\infty}^{\infty} dr\, \varphi_{\nu'}(r) V(r), \tag{C.4}$$

where $r = r_1 - r_2$, with

$$c_\nu = \int_{-\infty}^{\infty} dR\, \varphi_\nu(R) = 2^\nu \pi^{-1/4} \sqrt{2\nu! b} \sum_{\kappa=0}^{[\frac{\nu}{2}]} \left(-\frac{1}{8}\right)^\kappa \frac{\Gamma\left(\frac{\nu-2\kappa+1}{2}\right)}{\kappa!(\nu-2\kappa)!}, \tag{C.5}$$

where $R = r_1 + r_2$. If $V(r)$ has the Gaussian form given by Eq. (61), then

$$\int_{-\infty}^{\infty} dr\, \varphi_{\nu'}(r) V(r) = V_g 2^\nu \pi^{-1/4} \sqrt{2\nu! b} \sum_{\kappa=0}^{[\frac{\nu}{2}]} \left(-\frac{1}{8}\right)^\kappa \frac{\Gamma\left(\frac{\nu-2\kappa+1}{2}\right)}{\kappa!(\nu-2\kappa)!} \left(1 + \frac{2b^2}{a^2}\right)^{-\frac{\nu-2\kappa+1}{2}}. \tag{C.6}$$

Therefore the $b$-dependence of $\langle m_1 m_2 | \hat{V} | m_3 m_4 \rangle$ (and its corresponding $s$-dependence, arising from $s = b^{-1/2}$) is determined by terms containing

$$\varphi_m(0)\varphi_{m'}(0) b \left(1 + \frac{2b^2}{a^2}\right)^{-\frac{\nu-2\kappa+1}{2}} \sim \left(1 + \frac{2}{sa^2}\right)^{-\frac{\nu-2\kappa+1}{2}}. \tag{C.7}$$

The contribution from these terms into the hyperradial potential is obtained by applying the Laplace transform from Eq. (18), which gives

$$\frac{\sqrt{\Gamma(K+\frac{n}{2})\Gamma(K'+\frac{n}{2})}}{\rho^{K+K'+n-2}} \frac{1}{2\pi i} \int_{-i\infty}^{i\infty} ds\, e^{s\rho^2} s^{-\frac{K+K'+n}{2}} \left(1+\frac{2}{sa^2}\right)^{-\frac{\nu-2\kappa+1}{2}}$$

$$= \frac{\sqrt{\Gamma(K+\frac{n}{2})\Gamma(K'+\frac{n}{2})}}{\Gamma(\frac{K+K'+n}{2})} {}_1F_1\left(\frac{\nu-2\kappa+1}{2}, \frac{K+K'+n}{2}, -\frac{2\rho^2}{a^2}\right), \tag{C.8}$$

where ${}_1F_1$ is the confluent hypergeometric function. At very small values of the range $a$ the hypergeometrical function will the dominated by the first term of its asymptotic expansion, proportional to $1/\rho$, which is exactly the same as the hyperradial potential of the contact interaction.

## D Evaluation of Solid Angle

The solid angle appearing in Sect. 4.1 can be calculated by integrating a Gaussian function. In individual coordinates, one has

$$\int_{-\infty}^{+\infty} dr_1 \int_{-\infty}^{+\infty} dr_2 \ldots \int_{-\infty}^{+\infty} dr_N\, e^{-(r_1^2+r_2^2+\ldots r_N^2)} = (\sqrt{\pi})^N. \tag{D.1}$$

With (2), the same integral is also evaluated in hyperspherical coordinates by

$$\int_{-\infty}^{+\infty} dR \int_0^\infty \rho^{N-2} d\rho \int d\Omega_{N-1}\, e^{-(R^2+\rho^2)} = \frac{\sqrt{\pi}}{2} \Gamma\left(\frac{N-1}{2}\right) \int d\Omega_{N-1}. \tag{D.2}$$

Hence, one has

$$\int d\Omega_{N-1} = \frac{2(\sqrt{\pi})^{N-1}}{\Gamma(\frac{N-1}{2})}. \tag{D.3}$$

**Acknowledgements** N.K.T. is grateful for support from the United Kingdom Science and Technology Facilities Council (STFC) under Grant No. ST/L005743/1.

## References

1. Yu.F. Smirnov, K.V. Shitikova, The method of K harmonics and the shell model. Soviet Journal of Particles and Nuclei **8**, 344 (1977)
2. R.D. Amado, H.T. Coelho, $K$ harmonics in one dimension. Am. J. Phys. **46**, 1057 (1978)
3. W.G. Gibson, S.Y. Larsen, J. Popiel, Hyperspherical harmonics in one dimension: Adiabatic effective potentials for three particles with delta-function interactions. Phys. Rev. A **35**, 4919 (1987)
4. M. Gattobigio, A. Kievsky, M. Viviani, Spectra of helium clusters with up to six atoms using soft-core potentials. Phys. Rev. A **84**, 052503 (2011)
5. N.K. Timofeyuk, Convergence of the hyperspherical-harmonics expansion with increasing number of particles for bosonic systems. Phys. Rev. A **86**, 032507 (2012)
6. N.K. Timofeyuk, Shell model approach to construction of a hyperspherical basis for A identical particles: Application to hydrogen and helium isotopes. Phys. Rev. C **65**, 064306 (2002)
7. N.K. Timofeyuk, Improved procedure to construct a hyperspherical basis for the N-body problem: Application to bosonic systems. Phys. Rev. C **78**, 054314 (2008)
8. D. Baye, P.-H. Heenen, Generalised meshes for quantum mechanical problems. J. Phys. A **19**, 2041 (1986)
9. M. Vincke, L. Malegat, D. Baye, Regularization of singularities in Lagrange-mesh calculations. J. Phys. B **26**, 811 (1993)
10. D. Baye, The Lagrange-mesh method. Phys. Rep. **565**, 1 (2015)
11. D. Baye, M. Hesse, M. Vincke, The unexplained accuracy of the Lagrange-mesh method. Phys. Rev. E **65**, 026701 (2002)

12. P. Descouvemont, C. Daniel, D. Baye, Three-body systems with Lagrange-mesh techniques in hyperspherical coordinates. Phys. Rev. C **67**, 044309 (2003)
13. H. Bateman, *Tables of integral transforms*, vol. I (McGraw-Hill Book Company, New York, 1954)
14. M. Abramowitz, I.A. Stegun, *Handbook of mathematical functions* (Dover, New York, 1965)
15. D. Baye, Constant-step Lagrange meshes for central potentials. J. Phys. B **28**, 4399 (1995)
16. L. Filippin, M. Godefroid, D. Baye, Relativistic two-photon decay rates with the Lagrange-mesh method. Phys. Rev. A **93**, 012517 (2016)
17. D. Baye, Integrals of Lagrange functions and sum rules. J. Phys. A **44**, 395204 (2011)
18. J.B. McGuire, Study of exactly soluble one-dimensional N-body problem. J. Math. Phys. **5**, 622 (1964)
19. J. Avery, *Hyperspherical harmonics and generalized Sturmians* (Kluwer Academics Publishers, Dordrecht, 2000)
20. Yu.F. Smirnov, Talmi transformation for particle with different masses. Nucl. Phys. **39**, 346 (1962)